\documentclass[acmsmall]{acmart}

\setcopyright{acmlicensed}
\acmJournal{PACMMOD}
\acmYear{2023} \acmVolume{1} \acmNumber{1} \acmArticle{67} \acmMonth{5} \acmPrice{15.00}\acmDOI{10.1145/3588921}
\received{July 2022}
\received[revised]{October 2022}
\received[accepted]{November 2022}

\usepackage[a-1b]{pdfx}
\usepackage{hyperref}
\usepackage{array}
\usepackage{amsmath}
\usepackage{algorithmic}
\usepackage[ruled,vlined]{algorithm2e}
\usepackage{pgfplots,pgfplotstable}  
\usepackage{booktabs}
\usepackage{balance}
\usepackage{multirow}
\usepackage{enumitem}
\usepackage{adjustbox}
\usepackage{subcaption}
\usepackage{tablefootnote}
\usepackage{fancybox}
\usepackage{soul}

\newcommand{\edit}[1]{\textcolor{black}{#1}}

\newcommand{\eat}[1]{}
\newcommand{\msft}{Microsoft\xspace}
\newcommand{\cosmos}{Cosmos\xspace}
\newcommand{\scope}{SCOPE\xspace}
\newcommand{\rationorm}{Ratio-normalization\xspace}
\newcommand{\deltanorm}{Delta-normalization\xspace}
\newcommand{\bothnorm}{Ratio and Delta-normalization\xspace}
\newcommand{\smallsection}[1]{\vspace{1mm}\noindent\textbf{#1.}}

\newcounter{enum}

\begin{document}
\title{Runtime Variation in Big Data Analytics}

\author{Yiwen Zhu}
\affiliation{%
 \institution{Microsoft}
 \country{USA}
}
\email{yiwzh@microsoft.com}

\author{Rathijit Sen}
\affiliation{%
 \institution{Microsoft}
 \country{USA}
}
\email{rathijit.sen@microsoft.com}

\author{Robert Horton}
\affiliation{%
 \institution{Microsoft}
 \country{USA}
}
\email{rhorton@microsoft.com}

\author{John Mark Agosta}
\affiliation{%
 \institution{Microsoft}
 \country{USA}
}
\email{john-mark.agosta@microsoft.com}
\renewcommand{\shortauthors}{Yiwen Zhu et al.} 

\begin{abstract}

The dynamic nature of resource allocation and runtime conditions on Cloud
can result in high variability in a job’s runtime across multiple iterations, leading to a poor experience.
Identifying the sources of such variation and being able to predict and adjust for them is crucial to cloud service providers to design reliable data processing pipelines, provision and allocate resources, adjust pricing services, meet SLOs and debug performance hazards.

In this paper, we analyze the runtime variation of millions of production \scope jobs on \cosmos, an exabyte-scale internal analytics platform at Microsoft.
We propose an innovative 2-step approach to predict job runtime distribution by characterizing typical distribution shapes combined with a classification model with an average accuracy of >96\%,
out-performing traditional regression models and better capturing long tails.
We examine factors such as job plan characteristics and inputs, resource allocation, physical cluster heterogeneity and utilization, and scheduling policies.

To the best of our knowledge, this is the first study on predicting categories of runtime distributions for enterprise analytics workloads at scale.
Furthermore, we examine how our methods can be used to analyze what-if scenarios, focusing on the impact of resource allocation, scheduling, and physical cluster provisioning decisions on a job’s runtime consistency and predictability.
\end{abstract}



\maketitle

\section{Introduction}
\label{sec:intro}

\begin{figure}[t]
	\centering
			\includegraphics[width=0.65\columnwidth]{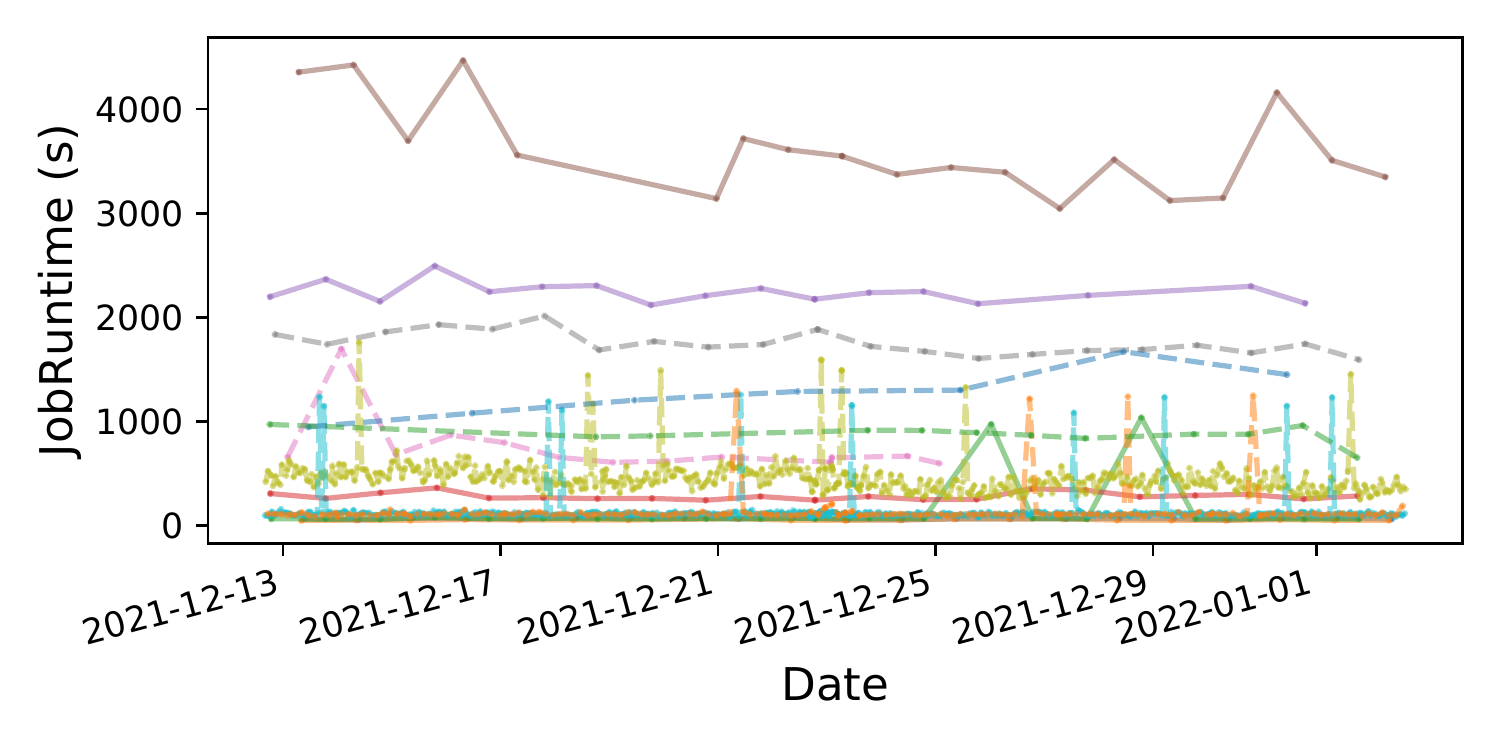}
			\vspace{-0.4cm}
\caption{\edit{Recurring jobs with runtime variation.}}
			\vspace{-0.4cm}
\label{fig:example-variation}
\end{figure}

Big Data platforms have become ubiquitous over the last decade, enabling scalable data processing with high efficiency, security, and usability~\cite{zaharia2010spark, thusoo2009hive,tigani2014google,chaiken2008scope,diaz2018azure,ramakrishnan2017azure,aws-athena, sarkar2018learning}. 
However, the dynamic nature of resource provisioning, scheduling, and co-location with other jobs can cause occasional job slowdowns. Additionally, intrinsic properties of the job such as parameter values and input data sizes can change across repeated runs leading to variations in runtime. Figure~\ref{fig:example-variation} shows a set of recurring jobs in \cosmos~\cite{power2021cosmos}, a Big Data analytics platform at Microsoft, submitted with different frequencies. We can see that some jobs have more stable runtime while some have occasional slow downs with non-regular patterns. But, it is not apparent \textit{why such variations are happening}, \textit{how they can be mitigated}, or \textit{how likely it is for the next job run to be an outlier} compared to historic runs.

In production systems, jobs are often scheduled or pipelined with strong data dependencies
(jobs using other jobs' output data as inputs)
~\cite{chung2020unearthing}.
Stability and predictability of job runtimes are important factors that affect the fundamental design and architecture of data processing pipelines.
Unfortunately, they are often neglected by operators due to the difficulties of assessment even though job slowdowns are inevitable~\cite{shao2019griffon}. 
Even with massive amounts of telemetry data, cloud providers still default to a manual triage process due to the difficulty of capturing the compounding factors that impact job runtime and its stability, which is not scalable and error-prone.

Although prior works\edit{~\cite{schad:ec2-variance:vldb:2010, feitelson1995job, zrigui2022improving} have empirically characterized runtime variation, they do not propose methods to predict the variation nor the likelihood of a new run being an outlier compared to the average or median runtimes.} Other works such as Griffon~\cite{shao2019griffon} used machine learning models to predict the minor slowdown in runtimes \edit{for a limited number of job templates. They are} unable to predict significant slowdowns that appear as outliers. As ML models are notoriously bad at handling outliers especially with a low existence, prior time-series based approaches~\cite{naghshnejad2018adaptive,sonmez2009trace} are not applicable. \edit{In this paper, we aim to address this gap for production data analytics systems by developing a novel and systematic approach for modeling, predicting, and explaining the job runtime variation, allowing for finer-grained differentiation in characteristics.}

For our study, we comprehensively examine the runtime variation for millions of production SCOPE~\cite{chaiken2008scope} jobs on Cosmos~\cite{power2021cosmos}, an exabyte-scale analytics platform at Microsoft that supports a broad spectrum of \msft products~\cite{power2021cosmos}. Our key contribution is a framework for systematically analyzing, predicting and explaining runtime variation that includes:
\begin{enumerate}[leftmargin=*]
 	\item \textit{Descriptive} analysis: by examining historic data including intrinsic job properties, resource allocation, and physical cluster conditions, we provide a better understanding of the factors affecting runtime variation for each individual job. In particular:
 	\begin{enumerate}[leftmargin=14pt]
 	    \item We show that popular scalar metrics, such as Coefficient of Variation (COV) considered by prior work~\cite{schad:ec2-variance:vldb:2010}, are not sufficient to characterize variation with the existence of outliers. Instead,  we propose a novel scheme of characterizing variation using properties of the distribution of normalized runtime of the jobs \edit{that provides fine-grained information 
 	such as the probability of outliers, quantiles, and shapes of the distribution.} [Section~\ref{sec:metrics}]
 	    \item We make novel use of Shapley values~\cite{shapley1953value} for explaining predictions for variation and quantitatively analyze the contributions of different features. [Section~\ref{sec:explain}]
 	\end{enumerate}
 	\item \textit{Predictive} analysis: we develop an innovative \edit{approach based on likelihood to identify distinctively-diverse runtime distributions, and predict the distribution with >96\% accuracy, out-performing previous methods~\cite{shao2019griffon}}. [Section~\ref{sec:model}]
 	\item \textit{Prescriptive} analysis: based on the predictor, we quantitatively analyze what-if scenarios and identify potential opportunities to reduce variation by limiting spare tokens, scheduling on newer generations of machines, and better load balancing. [Section~\ref{sec:prescriptive}]
\end{enumerate}



The rest of this paper is organized as follows. Section~\ref{sec:challenges} discusses challenges in estimating and predicting runtime variation and our goals and approach in this work. Section~\ref{sec:bg} gives a brief overview of \scope jobs in \cosmos, potential sources of variation, and the datasets that we study. Sections~\ref{sec:metrics}--\ref{sec:prescriptive} present the descriptive, predictive and prescriptive analyses as outlined above. Section~\ref{sec:related} discusses related work in this space and Section~\ref{sec:conclusion} concludes the paper.



\section{Goals, Challenges, and Approach}
\label{sec:challenges}

Reasoning about performance changes is often done manually by experienced engineers with strong assumptions that can potentially lead to biased results. More recently, the availability of massive telemetry data in the cloud, that includes both information about job characteristics as well as status of the physical clusters, and the advent of data analytic methods, raise expectations that this process can be improved with more systematic and rigorous approaches.


Our goal is to evaluate and predict runtime variation 
at the \textit{individual} job level
with a customized and use-case specific measurement that is more insightful for customers for both monitoring and planning purposes. And this is a highly desired metric from the customer's point of view, as validated by several conversations with the program managers (PMs). 
We also want to provide rich information regarding variation, such as the probability that a job runtime may exceed an extreme value, or various quantitative properties of the runtime distributions, e.g., quantiles, outliers, to the user, which was difficult to capture using traditional ML methods.

Moreover, performance modeling of computational jobs in distributed systems is difficult, especially when focusing on reliability, due to the following challenges.
\begin{itemize}[leftmargin=*]
    \item \textbf{Complex environmental factors (C1)}: Resource sharing in cloud computing platforms adds complexity to the modeling of the job runtime due to noisy neighbors and other environmental changing factors. It is untractable for manual approaches to relay the dynamic condition of each computation node and unravel the potential issues that result in performance degradation.

    \item \textbf{Existence of rare events (C2)}: For rare events (such as occasional service disruption) that result in outliers and longer tails of the runtime distributions, it is difficult to collect sufficient observations of outliers for a recurring job in order to accurately estimate their distributions. It is therefore crucial to be able to leverage the learning from job instances in other job groups that have sufficient observation samples. 
    \item \textbf{Lack of proper metrics (C3)}: How to measure variation remains a challenge in the case of the characteristic long-tailed distributions of runtime, for which conventional variance-based measures do not capture the extreme values of interest.
    Metrics such as COV that are commonly used to evaluate the runtime variation are not sufficient to capture detailed characteristics of various runtime distributions.
    \item \textbf{Lack of labeled data (C4)}: \edit{While the majority of machine learning approaches for predictive analysis require labeled data, there is no label recorded for the causes of runtime distributions or job slowdowns. Manually evaluating runtime reliability to determine the distribution \textit{category} each job belongs to is also infeasible for new jobs with a small number of occurrences. 
    }
\end{itemize}

\begin{figure}[h]
	\centering
	\vspace{-0.2cm}
			\includegraphics[width=0.7\columnwidth]{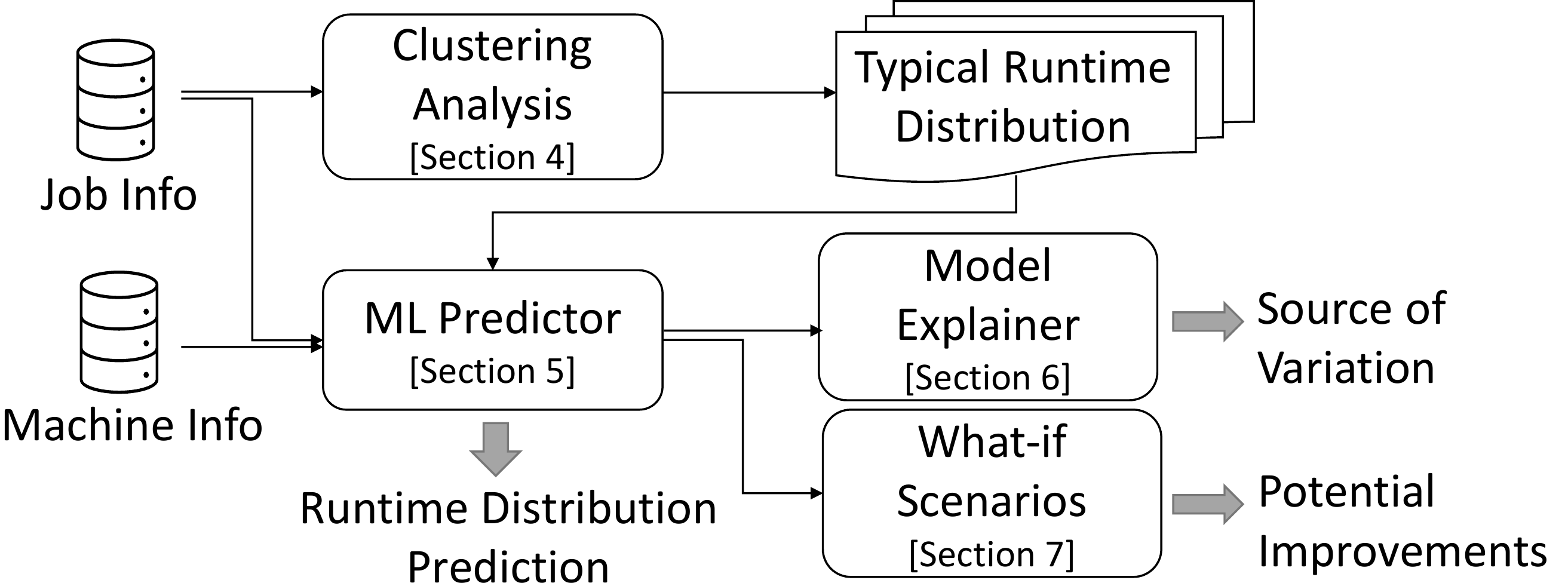}\vspace{-5pt}
\caption{\edit{Framework.}}
\label{fig:architecture}
\vspace{-8pt}
\end{figure}


Our 2-step approach in this work is to characterize and then predict variation based on the distribution of normalized runtimes of recurring jobs (see Figure~\ref{fig:architecture}). 
Leveraging information both at the job level and the machine level, we:

\begin{itemize}[leftmargin=*]
	\item \textbf{Characterize runtime distributions}: \edit{Our \textit{Clustering Analysis} uses a novel scheme of featurization to cluster~\cite{bishop2006pattern} historic jobs with distinctively-diverse runtime distributions. }
	We associate the job with one distribution it belongs to using an innovative and adaptive \textit{posterior likelihood} method. For each type (single-mode, multi-mode) of distribution, we define key metrics to depict the distribution, and quantify the variation in numeric terms that can be easily understood by users [challenge \textbf{C3}, Section~\ref{sec:metrics}].
	\item \textbf{Develop \textit{ML predictor}}: We leverage Machine Learning (ML) classification techniques to predict which distribution of runtimes the job most likely belongs to, taking into account job properties, resource allocation, and environmental conditions such as system load \edit{that potentially leads to noisy neighbors} [\textbf{C1}]. To overcome the challenges \textbf{C2} and \textbf{C4}, 
	we develop the model using the observations of distributions over a long time interval and for jobs with more recurrences (Dataset D1, Table~\ref{tab:datasets}), while the model can be applied to any new jobs. [Section~\ref{sec:model}]
	\item \textbf{Explain predictions}: We use \textit{Model Explainer} based on feature contribution algorithms to better understand the various factors associated with runtime variation [challenge \textbf{C1}, Section~\ref{sec:explain}]. 
	\item \textbf{Analyze \textit{what-if scenarios}}: Based on the prediction model, we propose hypothetical scenarios and evaluate the potential improvement of runtime performance quantitatively [Section~\ref{sec:prescriptive}].  
\end{itemize}

While point prediction for the job runtime is an important and challenging problem on its own\edit{~\cite{phoebe,pietri2014performance, tasq-arxiv, tasq-edbt, hu:reloca:infocom:2020, autodop,  starfish:analytics-cluster-sizing:socc:2011, rajan2016perforator}}, we want to predict the potential variation in runtimes for recurring jobs, rather than the absolute runtimes. Thus, direct prediction of job runtimes is a \textit{non-goal} for this work. 



\section{Platform and Datasets}
\label{sec:bg}

\cosmos~\cite{power2021cosmos} is an exabyte-scale big data platform developed at \msft since 2002, with more than 300k machines across multiple data centers worldwide~\cite{zhu2021kea}. Using a YARN-based~\cite{vavilapalli2013apache} resource manager, the system processes >600k jobs per day from tens of thousands of \msft internal users. It is a big internal shared-cluster where efficiency is paramount. Over the past decades, a multitude of research projects and engineering efforts have improved its efficiency, security, scalability and reliability~\cite{jindal2019peregrine, zhang2012optimizing, bruno2012recurring, bruno2013continuous,boutin2014apollo,boutin2015jetscope,karanasos2015mercury,jyothi2016morpheus,curino2019hydra,qiao2019hyper,chaiken2008scope,ferguson2012jockey,ramakrishnan2017azure,zhou2012advanced,zhu2021kea,zhou2010incorporating,wing2020osdi,phoebe,sen2021autoexecutor, zhu2021kea}. 

\cosmos jobs can be authored using a SQL-like dialect, named SCOPE~\cite{boutin2015jetscope} with heavy use of C\verb|#| and user-defined functions (UDFs). Upon submission, a job is compiled to an optimized execution plan as a DAG of operators, and distributed across different machines. 
Each job consists of multiple \textit{Vertices}, i.e., an individual process that will be executed on a container assigned to one physical machine.

\subsection{Job Groups}
Our work focuses on understanding and predicting variation in runtimes over repeated runs of jobs that we assemble into \textit{job groups}. Variation is meaningful only when jobs recur (i.e., sample size $>$ 1). Prior studies~\cite{jyothi2016morpheus,wing2020osdi,phoebe} have shown that 40--60\% of jobs on Cosmos are recurring jobs. Others~\cite{feitelson1995job} have also reported a significant fraction of jobs as recurring on their systems. 
We identify recurrences by matching on a key that combines the following.
\begin{itemize}[leftmargin=*]
    \item The normalized job name, which has specific information like submission time and input dataset removed~\cite{phoebe,jindal2019peregrine}. 
    \item The job signature~\cite{jindal2019peregrine}, which is a hash value computed recursively over the DAG of operators in the compiled plan. The signature does not include job input parameters.
\end{itemize}
We thus have \textit{job groups} with \textit{job instances} belonging to each group, corresponding to recurrences of the job. Job instances have the same key value within each job group.

\subsection{Sources of Variation}
\label{subsec:sources-of-variation}

Within each job group, runtimes of job instances can vary due to any of the following reasons:

\smallsection{Intrinsic \edit{characteristics}} The key used for grouping jobs includes information on the execution plan (e.g., type of operators, estimated cardinality, dependency between operators) while not including the job input parameters (e.g., parameters for filter predicates) or input datasets. Different instances can have different values for these parameters, datasets, and their sizes. This can lead to different runtimes within the group if the parameter changes are not accompanied by a change in the compiled plan. In our datasets, we have observed that input data sizes can vary by up to a factor of 50 within the same job group.
\begin{figure}
	\centering
		\includegraphics[width=0.65\columnwidth]{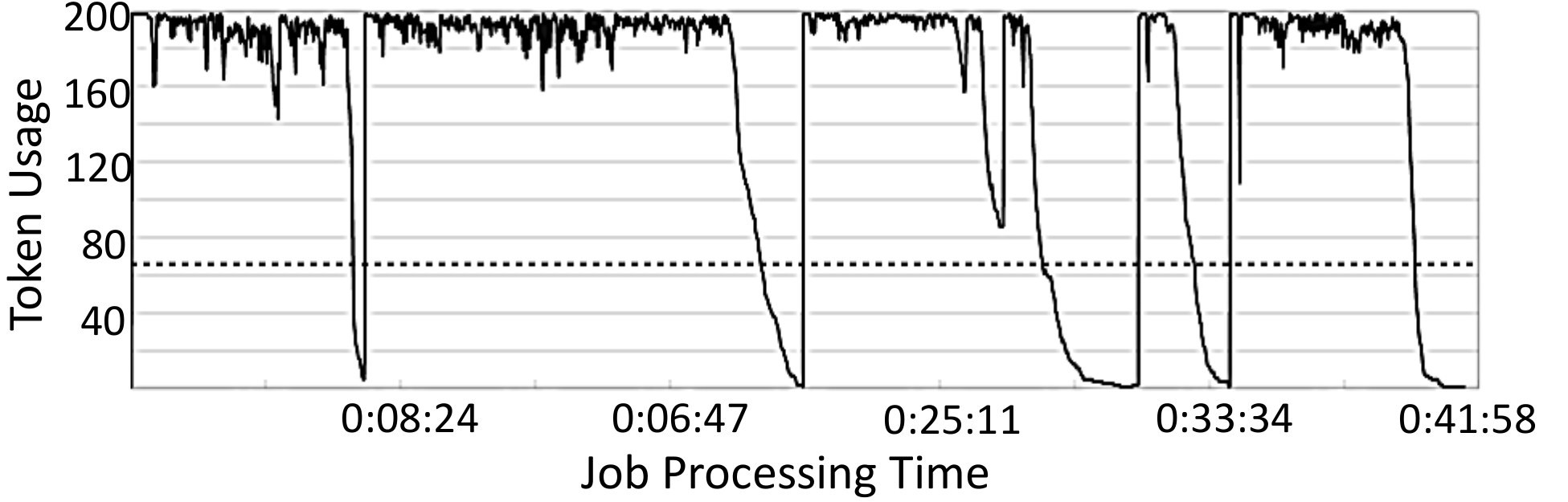}
	\vspace{-0.3cm}
	\caption{\edit{Token usage for an example job during its run.}
		\vspace{-0.4cm}
	}\label{fig:token}
\end{figure}

\smallsection{Resource \edit{allocation}} 
In \cosmos, the unit of resource allocation is a \textit{token}~\cite{power2021cosmos}, which is analogous to the notion of a container. 
The number of tokens guaranteed for a job can be specified by users at the time of job submission or it may be recommended by the system~\cite{autotoken}. To further improve the utilization of the existing infrastructure, unused resources are repurposed as preemptive spare tokens~\cite{boutin2014apollo} that can be leveraged by any jobs freely\footnote{The usage of spare tokens is capped by the allocation as specified by users.}. The availability of these spare tokens is difficult to predict, and can have meaningful effects on runtimes. Figure~\ref{fig:token} shows the skyline of token usage for a \cosmos job that was allocated with 66 tokens (dashed line). Including spare tokens, the job consumed up to 198 tokens in total throughout its processing time. 

The maximum number of tokens used by a job depends on how much parallelism it can exploit subject to the number of tokens allocated. While observing the execution of various workloads on \cosmos, we have seen maximum token counts vary by a factor of 10 within the same job group.
%
%
There is also variation in the characteristics of allocated resources. Tokens map to computational resources on compute nodes with different Stock Keeping Units (SKUs). Having evolved for over a decade, the \cosmos cluster consists of 10--20 different SKUs with different processing speeds~\cite{zhu2021kea}. In our datasets, we have observed different job instances within the same job group run on one to nine different SKUs simultaneously.

\smallsection{Physical cluster environment} Finally, physical cluster environment also leads to variations in runtimes. This includes both the availability of spare tokens (discussed above) and the load on the individual machines. 
Higher utilization (load) is likely to cause more contention for shared resources, and a larger range of loads may increase runtime variation.

\subsection{Datasets}

To develop insights into the sources of variation as discussed above, we collected data by:
i) extracting information about intrinsic characteristics such as operator counts in the plan, input data sizes, and cardinalities, costs, etc., estimated by the SCOPE optimizer using the Peregrine framework~\cite{jindal2019peregrine}; ii) obtaining token usage information from the job execution logs, and SKU and machine load information using the KEA framework~\cite{zhu2021kea}; and
iii) joining all this information together by matching on the job ID, name of the machine that executes each vertex, and the corresponding vertex start/end time.

\begin{table}[ht]
\centering
\vspace{-0.2cm}
\caption{Datasets used for this study. 
}\vspace{-0.26cm}
\label{tab:datasets}
\resizebox{0.6\columnwidth}{!}{\begin{tabular}{|c|c|c|c|c|}\hline
     \textbf{Dataset}&  \textbf{Interval}& \textbf{Job Groups}& \textbf{Job Instances}& \textbf{Support}\\\hline\hline
     \textbf{D1}& 6 months& >9K& >3M& 20\\\hline
     \textbf{D2}& 15 days& >11K& >700K& 3\\\hline
     \textbf{D3}& 5 days& >11K& >200K& 3\\\hline
\end{tabular}}
\end{table}

Table~\ref{tab:datasets} summarizes the datasets that we use for our study. The datasets consist of a subset of jobs run over the corresponding interval, and are included if the number of instances per group (support) exceeds a minimum threshold. With a support of minimum 3 occurrences, 53\% of jobs are included. 
In this research, we only focus on batch jobs as opposed to streaming jobs or interactive jobs that \cosmos also supports. 
We use dataset D1 to identify and group distributions of runtimes (see Section~\ref{sec:clustering}) for jobs with a large number of occurrences (>20).
We used D2 to train a predictor for runtime variation and D3 to test its accuracy (Section~\ref{sec:model}).


\section{Characterizing Runtime Variation}
\label{sec:metrics}

We now discuss how we characterize and quantify runtime variation for recurring jobs. This will form the basis for our prediction strategy that we discuss in Section~\ref{sec:model}.

\subsection{Scalar Metrics}

As outlined previously (Section~\ref{sec:intro}), a job's average runtime does not give much insight into variations across repeated runs, or how long the next run of the job will take. Thus, it is not very useful for characterizing, predicting, or explaining the variations.

\begin{figure}
	\centering
		\subfloat[\edit{Median vs instance runtimes}]{\includegraphics[width=0.30\columnwidth]{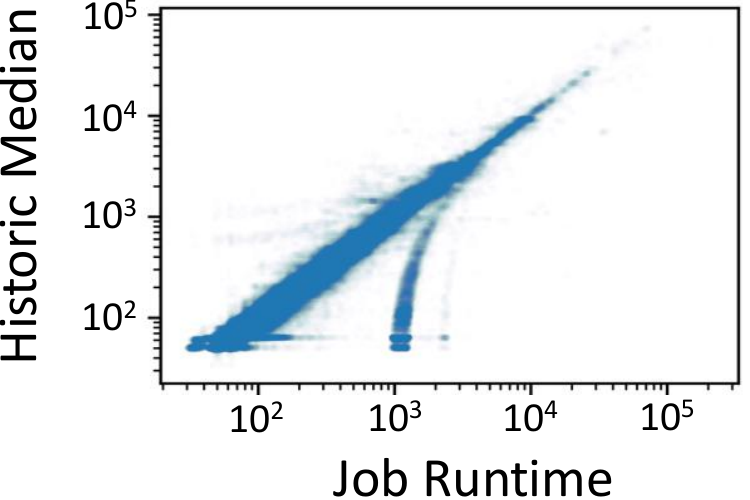}\label{fig:median-vs-runtime}\vspace{-0.1cm}}\quad\quad\quad
		\subfloat[\edit{Historic COV vs COV}]{\includegraphics[width=0.30\columnwidth]{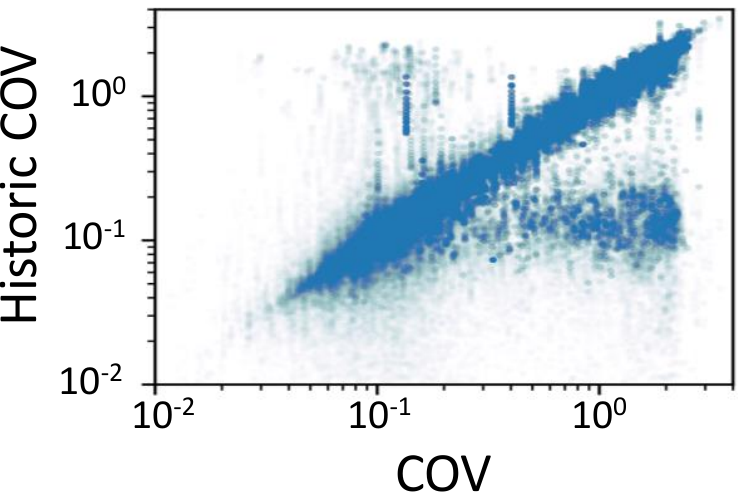}\label{fig:cov-as-predictor}\vspace{-0.1cm}}
		\vspace{-0.1cm}
	\caption{\edit{Correlations between historic median and job runtimes (a), and historic COV and COV of all observations (b).}}
	\label{fig:medianbenchmark}
			\vspace{-0.3cm}
\end{figure}
Next, we investigate how well a job's median runtime correlates with runtimes over the different repetitions of the job. Figure~\ref{fig:median-vs-runtime} shows how runtimes for individual repetitions of the job compare with its historic median using dataset D2 in log-scale. 

We observe two distinct patterns in Figure~\ref{fig:median-vs-runtime}---a set of points are clustered along the diagonal, indicating a good correlation of \edit{individual job instances' runtimes} to the median, and another set of points are clustered separately in a pattern resembling a stalagmite\footnote{A stalagmite is a rocky formation that arises from the floor of a cave and may reach the ceiling~\cite{stalagmite}.}. The runs corresponding to the points in the stalagmite are much slower than the median runtime and contribute to the (long) tail of the runtime distributions. 
Such runs are rare (comprising less than 5\% of all runs), with the probability reducing with larger median values. But we have found it to be very difficult to predict upfront if the time for a new run will end up on the diagonal or on the stalagmite. The existence of these two patterns, as well as the difficulty of predicting to which pattern a new run will belong, even if the median runtime is stable and known, makes the median a poor choice to use for predicting runtimes or for characterizing variations. A similar trend can be found for the average and $95^{\text{th}}$ percentile of historic runtime. 



The Coefficient of Variation (COV) is another commonly-used metric to characterize variation. It is defined as the (unitless) ratio of standard deviation to the average. COV is straightforward to compute and interpret, and prior work~\cite{schad:ec2-variance:vldb:2010} has used COV to characterize variation in job runtimes.
%
%
But COV has several limitations:
\begin{itemize}[leftmargin=*]
    \item \textbf{Bias}: The runtimes of SCOPE jobs that we study range from seconds to days, with significant differences in the average values. This may cause COV to be biased and one could always observe very large COV for short-running jobs.
    \item \textbf{Instability}: The average runtime can increase due to the existence of outliers (in such large distributed systems, some jobs inevitably run slow occasionally). Thus, COV can be unstable with addition of more jobs in the dataset. Unlike the average, COV does not converge with a large sample size thus does not have a consistent estimator~\cite{fischer2011history}.
    \item \textbf{Coarse-grained}: COV does not capture many characteristics of a distribution, such as its shape (such as unimodal, bimodal, and the existence of outliers). Hence it cannot readily explain variation in a fine-grained manner.
\end{itemize}

\edit{Figure~\ref{fig:cov-as-predictor} shows how well the COV computed from historic runs (y-axis) for each job instance based on dataset D2 compares with the COV of times from all runs (x-axis) based on the observation in D3.} Similar to the discussion for medians above, we see multiple groups of points, with the same historic COV appearing for different COV values from new runs and it is difficult to predict for a new run to which group it will belong. Additionally, the COV metric suffers from the limitations mentioned above. 

Overall, we found that scalar metrics such as average, median, quantiles, and COV by themselves are not sufficient for understanding or predicting runtime variation.

\subsection{Distributions}
\label{sec:clustering}

We represent runtime variation for each recurring job group by its \textit{runtime distribution}. Although there is a large variety in the runtimes of SCOPE jobs, we found that runtimes of many different jobs have similar probability distributions. We refer to these as \textit{shapes}. Knowing a job's distribution is sufficient to determine any characteristic of its variation, including the risk that its runtime will exceed a specified threshold. 

To compute the shapes, we first normalize the job runtimes, then compute their empirical Probability Mass Functions (PMFs, i.e., histogram).
Jobs are \textit{clustered}
based on the similarity of their runtime distributions, and for any new job, we can predict the cluster it belongs to. We identify the job's PMF as that of the cluster it belongs to. This methodology allows us to generalize our analysis across different jobs and work instead with a small number of clusters that can be easily understood by the users.


We used the following two normalization strategies to transform job runtimes, using medians computed on ``historic'' data from Dataset D1 as in Table~\ref{tab:datasets}:

\begin{definition}
\textit{The} \textbf{\rationorm}\textit{ is defined by the ratio of job runtime to its historic median, i.e., job runtime / median runtime. And }\textbf{\deltanorm}\textit{ is defined by the difference, job runtime - median runtime.}
\end{definition}

The \rationorm distribution measures relative change in runtimes, while the \deltanorm distribution measures the absolute deviation from median, measured in seconds. \edit{Note that runtime with various ranges can be normalized more effectively using ratio-normalization. E.g., absolute variations for long-running jobs are typically higher (1h$\pm$10min), whereas those for short-running jobs can be a lot lower. In this regard, ratio-normalization improves consistency and lumps together comparable distributions with different runtime ranges.
On the other hand, for very short or very long jobs, it might be less insightful to measure variances in percentage. For short-running jobs, the percentage can be very large (e.g., 5s$\pm$300\%). For long jobs, the percentage can be very small, leading to a very "thin" distribution measured by the ratio-normalization. Therefore, in this work, we leverage the delta-normalization combined with ratio-normalization to capture the variation in absolute terms.}

Our clustering analysis to recover the ``typical'' distribution shape across jobs uses dataset D1 where only jobs with >20 occurrences are included for more accurate estimation for their runtime distribution. For each job group, we derive its histogram for the distribution of normalized runtimes and then use an unsupervised machine learning algorithm to cluster them. Note that the inputs to the clustering analysis are the PMF probabilities of each bin of the histogram as opposed to the job features (e.g., input size, etc.). 

Our principal design choices for the runtime distribution clustering method are as follows:
\begin{itemize}[leftmargin=*]
	\item \textbf{Bin size and Range}: 
	The range should cover the majority of values with relatively fine granularity but not too small to capture fluctuation due to noise. 
	We merge the outliers into one bin (based on being $\leq$ or $\geq$ some thresholds)\footnote{For \deltanorm, we use [-900, 900] (where 1\% of jobs are 1066s slower than median, we round down to 900s, i.e., 15 min), and for \rationorm, we use [0, 10] (where 1\% of jobs are 10.6x slower than median, we round down to 10x). Jobs >900s or 10x slower than median are defined as outliers.}.  
	We evaluated 50, 100, 200 and 500 bins, and chose 200 bins that has relatively smooth PMF curves, and different shapes of distributions are observable.
    \item \textbf{Clustering algorithm}: Hierarchy clustering based on dendrogram~\cite{nielsen2016introduction} and Agglomerative clustering~\cite{rokach2005clustering} take different distance metrics, linkage methods, and user-specified number of clusters. However, they result in imbalanced clusters (some with >90\% of the data in one cluster). K-means clustering~\cite{sculley2010web} resulted in more balanced clusters, so is chosen for the following analysis. 
	\item \textbf{Number of clusters}: It is determined based on: (i) numeric analysis of \textit{inertia}, defined by the sum of squared distances between each sample and its cluster centroid (we pick an elbow point where adding more clusters does not significantly decrease the inertia), 
	and (ii) by visually examining the clustering results to check if the clusters are sufficiently different from each other and have unique characteristics. 
	\item \textbf{Smoothing histograms}: The standard clustering algorithms are based on using PMF probabilities as input vectors assuming each dimension is independent. In reality, adjacent density values of bins (e.g., the probability of a runtime being in the $4^{\text{th}}$ or $5^{\text{th}}$ bin) are correlated. However, with any distance measurement (e.g., dot product), correlation between adjacent bins is not considered.
	Therefore, we introduce a \textit{smoothing step} after deriving the PMFs to jointly consider any adjacent bins' values such that the two smoothed vectors mentioned above will have a higher affinity. 
\end{itemize}


%

\begin{figure}[h]
\vspace{-0.2cm}
	\centering
	\subfloat[\edit{\rationorm}]{\includegraphics[width=0.7\columnwidth]{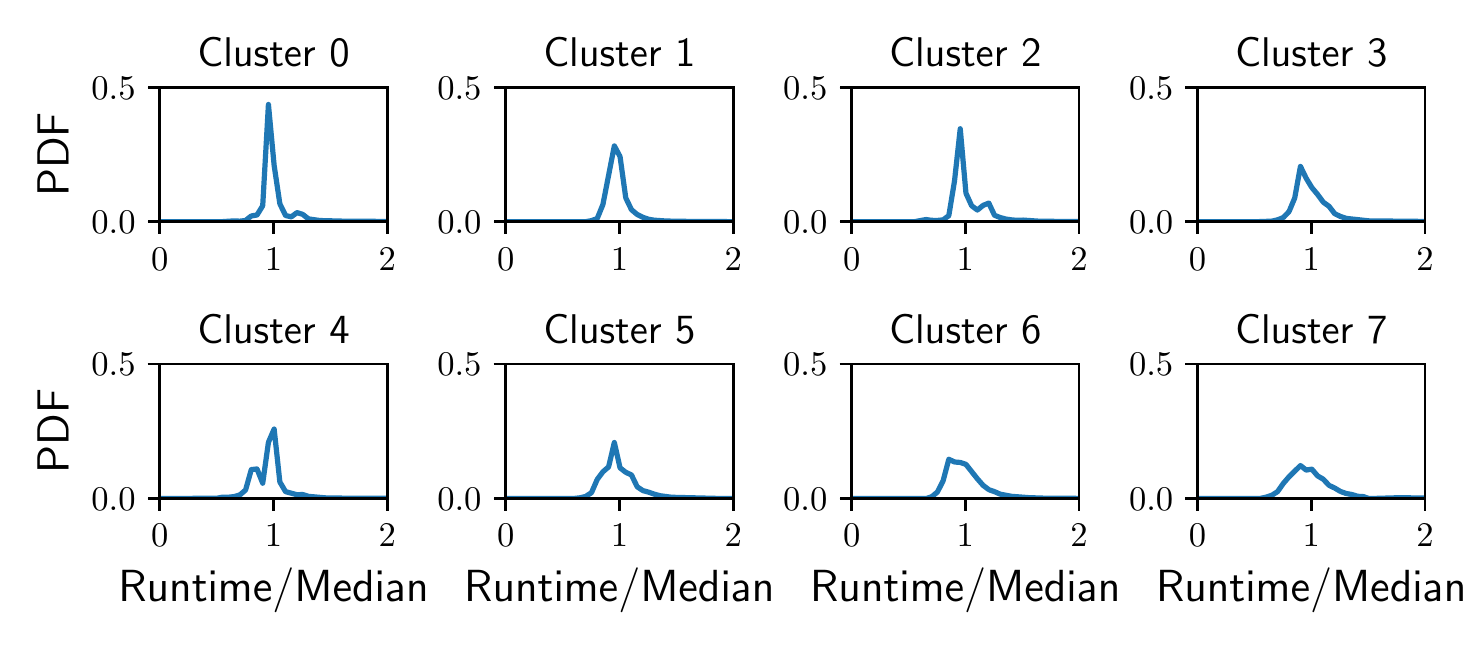}\vspace{-0.1cm}}\\
    \subfloat[\edit{\deltanorm}]{
		\includegraphics[width=0.7\columnwidth]{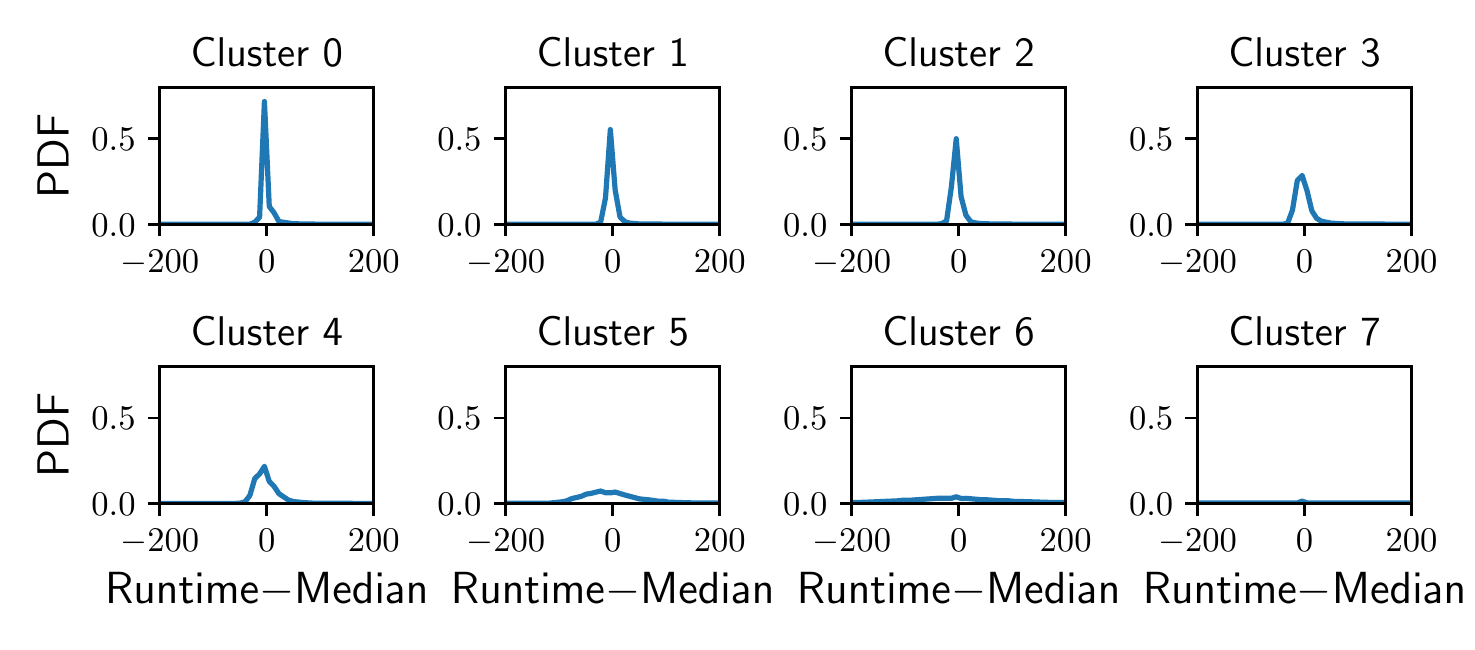}\vspace{-0.3cm}}
	 	\vspace{-1mm}
	\caption{\edit{Typical distributions of normalized runtime.
	}}\label{fig:clustering}
\end{figure}

Figure~\ref{fig:clustering} shows the distributions for the 8 clusters using \rationorm and \deltanorm policies. We see that some distributions have two modes (e.g., Cluster 0, 2, 4 using \rationorm) and with different variances. Table~\ref{tab:statscluster} summarizes important statistics for each cluster. 
For example, Cluster 0 with \rationorm has a outlier probability of 1.63\% (defined by $\geq 10$x slower than median for \rationorm);
the difference between 25 and $75^{\text{th}}$ percentile is 0.06; the $95^{\text{th}}$ percentile of this distribution is 1.41, and the standard deviation is 2.46. 
The outlier probability decreases to 0.06\% for Cluster 7 with \rationorm.
Clusters are ranked according to the difference between the $25^{\text{th}}$ and $75^{\text{th}}$ percentiles.
\begin{table}[h]
	\caption{Statistics for the clusters of runtime distributions.}\label{tab:statscluster}
\begin{adjustbox}{max width=0.67\columnwidth}
\begin{tabular}{lp{0.8cm}p{0.5cm}ll|lp{0.8cm}p{1.0cm}p{0.8cm}l}
		\toprule
\multicolumn{5}{c}{Ratio}             & \multicolumn{5}{c}{Delta}             \\\midrule
cid & outlier (\%) & $25-75^{\text{th}}$ & $95^{\text{th}}$ & std & cid & outlier (\%) & $25-75^{\text{th}}$ (s)& $95^{\text{th}}$ (s)& std (s)\\\midrule
0   &  1.63                           & 0.06 & 1.41                           & \textbf{2.46} & 0  & 1.93                            & 4   & 28                             & 155                            \\
1   &  0.42                           & 0.11 & 1.2                            & 0.93                           & 1  & 0.49                            & 11  & 19                             & 140                            \\
2   &  \textbf{1.66} & 0.16 & 1.37                           & 2.18                           & 2  & \edit{0.53}                            & 11  & 23                             & 148                            \\
3   &  0.25                           & 0.17 & 1.29                           & 1.45                           & 3  & 0.55                            & 16  & 33                             & 140                            \\
4   & 1.46                           & 0.17 & 1.35                           & 1.94                           & 4  & 0.98                            & 31  & 63                             & 153                            \\
5   & 0.25                           & 0.19 & 1.34                           & 0.82                           & 5  & 0.73                            & 69  & 128                            & 179                            \\
6   & 0.26                           & 0.20  & 1.37                           & 0.97                           & 6  & 2.43                            & 199 & 408                            & 296                            \\
7   & 0.06                           & \textbf{0.29} & \textbf{1.46} & 0.55                           & 7  & \textbf{24.23} & \textbf{936} & \textbf{1359} & \textbf{2548} \\\bottomrule
\end{tabular}
\end{adjustbox}
	\vspace{-0.4cm}
\end{table}

\section{Predicting Runtime Variation}
\label{sec:model}


We now develop a prediction model based on explainable ML to 
predict the \textit{shape} of runtime distribution (as in Figure~\ref{fig:clustering}) of jobs. We use dataset D2 as training set and D3 as testing set.

\subsection{Feature Selection}\label{sec:input}

We consider three classes of predictive features that are available at the job compile time: those derived from the job execution plan ("intrinsic"), those representing statistics of the job's past resource use, and features describing the load in the physical cluster where the job will run. We describe the classes below.

\smallsection{Intrinsic characteristics}
We leverage information on the job execution plan obtained from the query optimizer~\cite{jindal2019peregrine} at compile time as input, which can be indicative of the query type, data schema and its potential computation complexity. It includes the number of operators in each type (e.g., Extract, Filter), estimated cardinality, etc.. 
For a newly submitted job, its detailed input data size is unknown, and the estimated cardinality can be quite off~\cite{phoebe}. Therefore, using historic job instances of the same job group, we extract statistics for the total data read, temp data read, as well as the statistics related to the execution plan as additional input features that can be informative for the size of the job. 

We also derive the fraction of vertices running on each SKU as the input features, which indicates the resource consumption by each SKU. A previous study~\cite{zhu2021kea} shows that, in \cosmos, some newer SKUs might process data faster than the others; therefore, we believe that the \edit{fractions} of vertices executed on different SKUs would impact the runtime distribution. 

\smallsection{Resource allocation}
\edit{The token allocation is a good indicator for the resources being utilized by a particular job thus impacting the runtime. However, \cite{autotoken} detects that users are often over-allocating (e.g., user selects to allocate 1000 tokens, but the peak actual usage is only 600). In this work, we integrated historic token utilization with token allocation as the input. For historic job instances of the same job group, we extract the resource utilization (min, max, and average token usage based on the skyline as in Figure~\ref{fig:token}) and use the historic statistics as features (historic average and standard deviation). We also created a new variable for spare tokens (historic average). The model learned to place less importance on token allocation as a feature compared to actual utilization, and we corroborate this from the Shapley scores in Section~\ref{sec:exp}.}


\smallsection{Physical cluster environment}
The job runtime can be affected by the utilization of the machines that execute its vertices---a higher utilization level indicates a hotter machine that is likely to have more severe issues related to noisy neighbors and resource contention. Therefore, we extract the CPU utilization level of the corresponding machines in each SKU at the job submission time as the input. \edit{Compared to existing methods, such as~\cite{shao2019griffon} and~\cite{zrigui2022improving}, incorporating new-real-time machine status information improves the model accuracy (see Section~\ref{sec:model}).}

\edit{Work is ongoing to record the per container usage for both CPU and memory that includes more targeted information on the particular job and captures more accurately the resource consumption compared to the machine-level counters~\cite{cpuwrong}. Once available, features can easily be replaced or added to our models. We expect them to be strong indicators for the job runtime as they capture more job-level charateristics. They might also reveal if a job that is CPU intensive or memory intensive is more likely to have large runtime variances.}

\subsection{Cluster membership prediction}\label{sec:exp}
For a new job to be submitted, we want to \textit{predict} its runtime distribution shape based on information that is available at compile time. This naturally maps to a classification problem, where the prediction target is to map each job to a particular distribution shape (e.g., one among the 8 different shapes as in Figure~\ref{fig:clustering}).


\smallsection{Cluster membership based on posterior likelihood}
For each job instance, to determine which distribution shape it has, we leverage the set of similar job instances in the analyzed period with the same job group (i.e., same job name and execution plan) to derive the group's empirical Probability Mass Function (PMF), i.e., the histogram of the runtime distribution. 
Based on even a small number of runtime observations, we derive the posterior likelihood of these observations to be drawn from any one of the pre-defined distribution shapes as in Figure~\ref{fig:clustering}. 

Based on Bayes' Theorem~\cite{bishop2006pattern}, the \textit{posterior log-likelihood} that a job group with $N$ runtime observations, $x_{n=1 \cdots N}$, belongs to a cluster $z_{i=1 \cdots K}$ can be derived based on the PMF of these $N$ observations, $\phi_{h = 1 \cdots H}$, and the PMFs of the $K=8$ pre-defined clusters, $\theta^{i = 1 \cdots K}_{h = 1 \cdots H}$, which is adaptive to larger sample size:\vspace{-5pt} 

{\small
\begin{align}
p(x_1, x_2 \cdots x_N|z_i) &= \prod_{n=1 \cdots N} F(x_n|\theta^{i}_{h = 1 \cdots H}) \\
&= \prod_{n=1 \cdots N} \theta^{i}_{h(x_n)}\\
\log p(x_1, x_2 \cdots x_N|z_i) &= \sum_{n=1 \cdots N} \log (\theta^{i}_{h(x_n)})\\
p(z_i|x_1, x_2 \cdots x_N) &= \frac{ p(x_1, x_2 \cdots x_N|z_i)p(z_i)}{\sum_{i=1 \cdots K}p(x_1, x_2 \cdots x_N|z_i)p(z_i)} \\
&= \frac{ \prod_{n=1 \cdots N} \theta^{i}_{h(x_n)}}{\sum_{i=1 \cdots K}\prod_{n=1 \cdots N} \theta^{i}_{h(x_n)}}\\
&\sim \prod_{n=1 \cdots N} \theta^{i}_{h(x_n)}\\
\log p(z_i|x_1, x_2 \cdots x_N) &= \sum_{n=1 \cdots N} \log \left(\theta^{i}_{h(x_n)}\right) - \text{constant}\\
&= \sum_{h=1 \cdots H} n_h \log \left(\theta^{i}_{h}\right)- \text{constant}\\
&\sim \sum_{h=1 \cdots H} \phi_h  \log \left(\theta^{i}_{h}\right)\label{eq:loglikelihood}
\vspace{-60pt}
\end{align}
}Where,

\begin{tabular}[h]{lp{11.7cm}}
	$H$:& number of discrete bins when we derive the PMF for each distribution, a constant across all distributions.	\\
	$\theta^{i = 1 \cdots K}_{h = 1 \cdots H}$:& parameter of normalized runtime distribution for cluster $i$, specifically, the PMF value for bin $h$.\\
	$\phi_{h = 1 \cdots H}$:& parameter of distribution based on observations for a particular job group (i.e., $x_{n=1,2 \cdots N}$), specifically, the probability for bin $h$ of the PMF.\\
	$h(x_n)$:& the bin index that observation $x_n$ belongs to.\\
	$n_{h}$:& number of observations of runtime (i.e., $x_{i=1 \cdots N}$) for the job group that belongs to bin $h$.\\
	$x_{n=1 \cdots N}$:& runtime observation $n$, where
	$x_{n=1 \cdots N}|z_{i=1 \cdots K} \sim F(\theta^{i = 1 \cdots K}_{h = 1 \cdots H})$.\\
%
	$p(z_i)$:& prior on the probability of each cluster, assuming to be a constant across all clusters (non-informative prior~\cite{bishop2006pattern}).\\
\end{tabular}

It is interesting to point out that the log-posterior-likelihood is proportional to the dot product of the PMF of observations for the particular job group, i.e. $\phi_{h}$,  and the one of the pre-defined 8 clusters (after taking the log), i.e., $\theta^{i}_{h}$ (see Equation~\ref{eq:loglikelihood}).

\begin{figure}
	\centering
	\subfloat[\edit{Most likely cluster\label{fig:ll1}}]{			\includegraphics[width=.31\columnwidth]{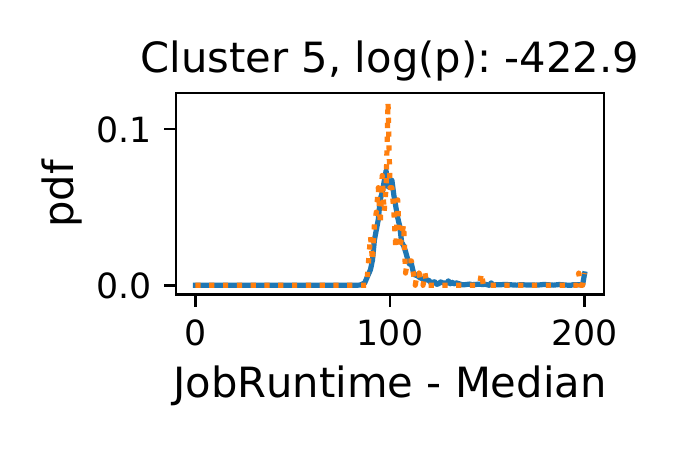}\vspace{-8pt}}\quad\quad
	 \subfloat[\edit{Least likely cluster\label{fig:ll2}}]{			\includegraphics[width=.31\columnwidth]{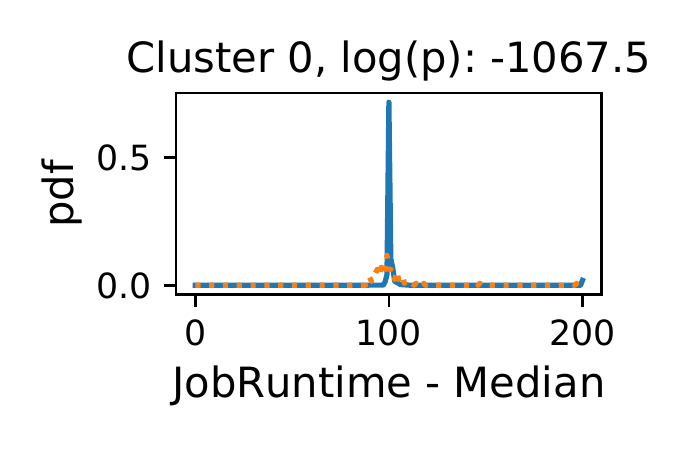}\vspace{-8pt}}
	 \vspace{-8pt}
	\caption{\edit{Examples of likelihood values (higher values indicate more probable).}}\label{fig:ll}
	\vspace{-0.5cm}
\end{figure}

Figure~\ref{fig:ll} shows an example of normalized runtime distribution (by \deltanorm) for a job with 10 occurrences compared with 2 clusters. The dashed line is the PMF for observations for this job group, i.e. $\phi_{h}$, and the solid line is for the predefined clusters, $\theta^{i}_{h}$. We see that Cluster 5, with the highest log-likelihood of -422.9, has the most similar shape \edit{(see Figure~\ref{fig:ll1}), while Cluster 0 has the least similar shape (see Figure~\ref{fig:ll2}).} Each job instance together with its job group is then associated with a cluster label with the highest likelihood as the prediction target (label).
\edit{This cluster association algorithm will always place a job in the most likely cluster. We observed that jobs with fewer observations may not be similar to any existing cluster’s runtime distribution. Therefore, we focus only on job groups with sufficient samples. We employ the inertia curve to tune the number of clusters as proposed in Section~\ref{sec:clustering} to avoid overfitting.
}



\smallsection{Classification model}
Based on the inputs, we conduct (1) passive-aggressive feature selection~\cite{zheng2015online} based on feature importance to avoid the use of correlated features, (2) parameter sweeping to select the best hyper-parameters for the classification algorithm, such as the number of trees for tree-based algorithms, and (3) fitting using \edit{RandomForestClassifier~\cite{rf}, LightGBMClassifier~\cite{lgbm} and EnsembledClassifier~\cite{es} by combining a set of popular classification algorithms, such as RandomForestClassifier, LightGBMClassifier, GradientBoostingClassifier~\cite{gb}, GaussianNB~\cite{gn}, and XGBClassifier~\cite{xgb}, using soft voting.} Note that RandomForestClassifier and LightGBMClassifier are well-known to have high accuracy for ML tasks using tabular data\edit{, especially for out-of-sample tests}. \edit{In this work, among the classifiers,} LightGBMClassifier has the highest accuracy\edit{, thus we report its result for the rest of the paper. By analyzing the prediction results, we noticed:}

    

By examining the Gini importance~\cite{louppe2013understanding} of the input features, we found that features related to the computation complexity and input data sizes (such as count of vertices, and data read) are significant and the features related to the historic runtime observations are also significant.
The token utilization (such as the max), and compile time information (such as cardinality estimates) are also important. The CPU utilization of machines
also impacts the prediction, which coincides with our belief that the physical cluster environment will affect the runtime variation of jobs. 
\edit{We also noticed that many of the operators turned out to be less important. The total vertex count is less important than the data size (total data read or cardinality-related metrics). It is possibly due to the huge variation in data processed by each vertex.}
In Section~\ref{sec:explain}, we dive into more details on the contribution of some features.
\begin{figure}
	\centering
    \subfloat[Confusion matrix\label{fig:cm}]{			\includegraphics[width=.43\columnwidth]{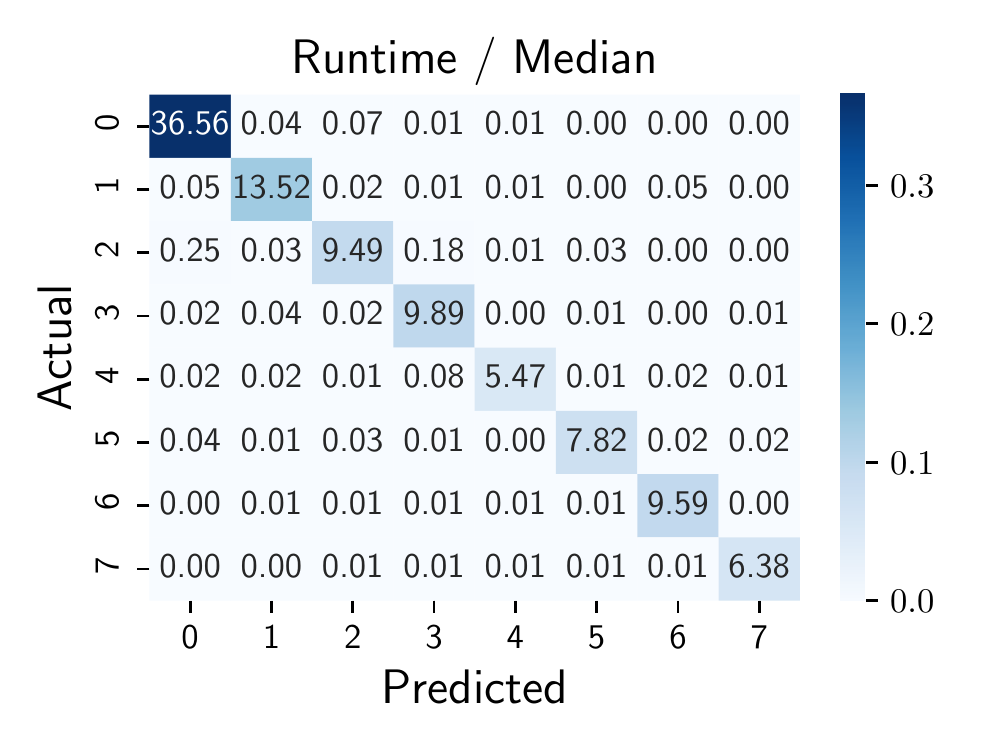}\vspace{-8pt}}
        \subfloat[Accuracy by occurrences\label{fig:accuracy}]{			\includegraphics[width=.42\columnwidth]{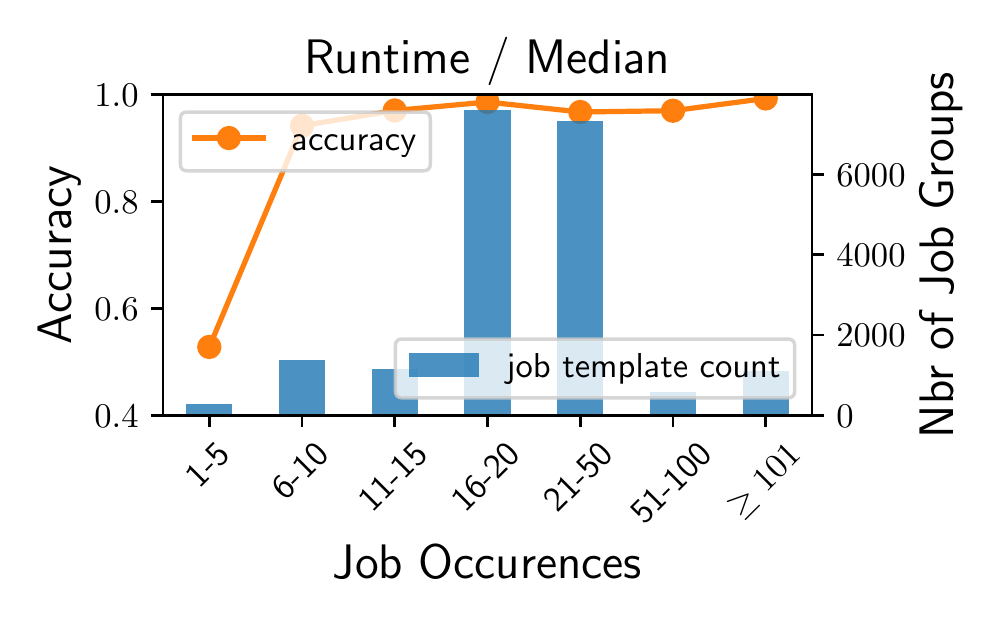}\vspace{-8pt}}
    \vspace{-3mm}
	\caption{\edit{Prediction accuracy for \rationorm.}}\label{fig:cm2}
	\vspace{-10pt}
\end{figure} 
\begin{center}
\shadowbox{
\begin{minipage}{0.9\linewidth}
    \textbf{Insight:} \edit{The feature importance learned from the model is mostly consistent with our expectation.}
\end{minipage}
}\vspace{-3pt}
\end{center}


Figure~\ref{fig:cm} shows the confusion matrix on test data comparing the predicted label (the x-axis) and the actual label (the y-axis) where each cell shows the portion of jobs of each category.
Predictions using both \bothnorm achieve overall accuracy of >96\%.
Figure~\ref{fig:accuracy} (orange line) shows the accuracy for jobs with different numbers of historic occurrences. We can see that for jobs with more historic occurrences, the prediction accuracy is higher, which indicates that the model prediction can be further refined by adding more observations from the same job group. The blue bar shows the count of job groups based on the number of job occurrences (1-5, 6-10, etc.). We can see that most of the jobs have 16-50 historic observations over the analyzed period. 
Similar trend can be seen for \deltanorm.
\begin{center}
\shadowbox{
\begin{minipage}{0.9\linewidth}
    \textbf{Insight:} \edit{Model predictions using both \bothnorm achieve high accuracy.}
\end{minipage}
}\vspace{-3pt}
\end{center}

\begin{figure}
	\centering
    \subfloat[Regression\label{fig:compare_regre}]{
    \includegraphics[width=.42\columnwidth]{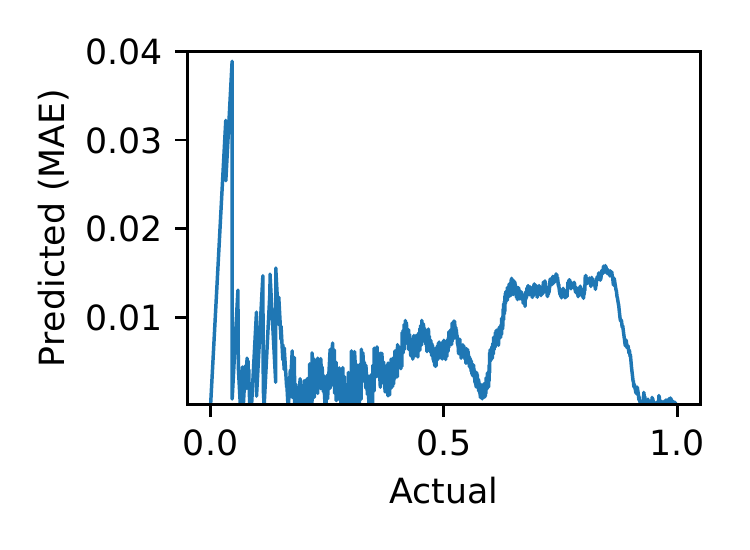}\vspace{-8pt}}
    \subfloat[Proposed approach\label{fig:compare_class}]{			
    \includegraphics[width=.42\columnwidth]{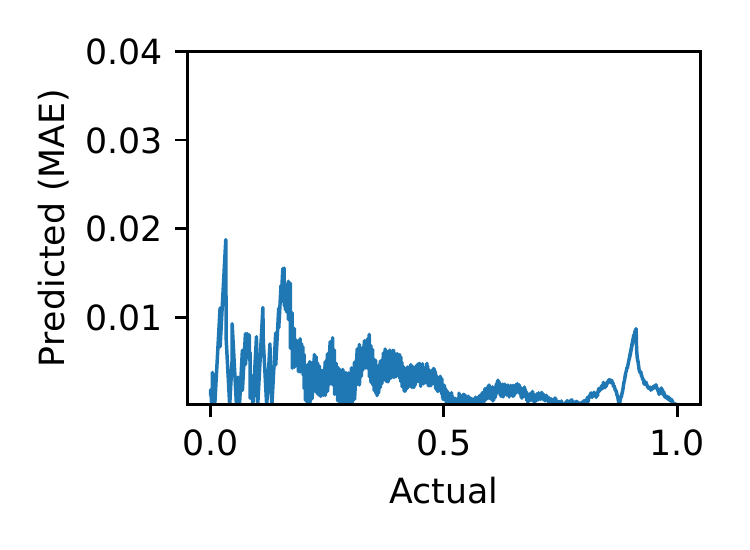}\vspace{-8pt}}
    \vspace{-2mm}
	\caption{Prediction accuracy for \deltanorm compared with traditional regression model.\label{fig:compare}}
	\vspace{-10pt}
\end{figure}

\edit{We extended the traditional random forest regression model as proposed in~\cite{shao2019griffon} by adding more query optimizer and near-real-time machine status information as features to predict the job runtime as the label}. Figure~\ref{fig:compare} compares the predicted distribution for all job runtimes based on the proposed method (using classification model to predict the distribution shape) and the regression model against the actual job runtime distribution using Quantile-Quantile Plot~\cite{gnanadesikan1968probability}, plotting the mean absolute error (MAE) in the y-axis. If two distributions are identical, the plotted quantile should align and the MAE=0. We can see that the proposed classification model (Figure~\ref{fig:compare_class}) has better accuracy compared with the traditional regression model (Figure~\ref{fig:compare_regre}) especially for higher percentiles as it captures better the existence of outliers in the distributions of clusters of jobs (see Figure~\ref{fig:clustering}). The Kolmogorov–Smirnov distance~\cite{karson1968handbook} is also reduced by 9.2\%, indicating better accuracy. 
\begin{center}
\shadowbox{
\begin{minipage}{0.9\linewidth}
    \textbf{Insight:} \edit{The proposed method outperforms existing model in predicting outliers.}
\end{minipage}
}\vspace{-3pt}
\end{center}



\section{Explaining Runtime Variation}
\label{sec:explain}

In this section, we conduct \textit{descriptive analysis} to better understand the job characteristics that lead to different runtime distributions. 
Starting with the classification models from the previous section, we use machine learning explanation tools to better understand the sources of runtime variation. 



\subsection{Shapley Value}

Shapley values~\cite{shapley1953value,lundberg2017unified} that explain the contribution of each ``player'' in a game-theoretic setting have been adopted for explaining the contribution of features in ML models. In our context, based on the predictors developed in Section~\ref{sec:model}, they explain the quantitative contribution of each feature by randomly permuting other feature values and evaluating the marginal changes of the predictions~\cite{molnar2020interpretable}. 

 \begin{figure}
	\centering
    \subfloat[By Input Size\label{fig:indi_contribution}]{			\includegraphics[width=.4\columnwidth]{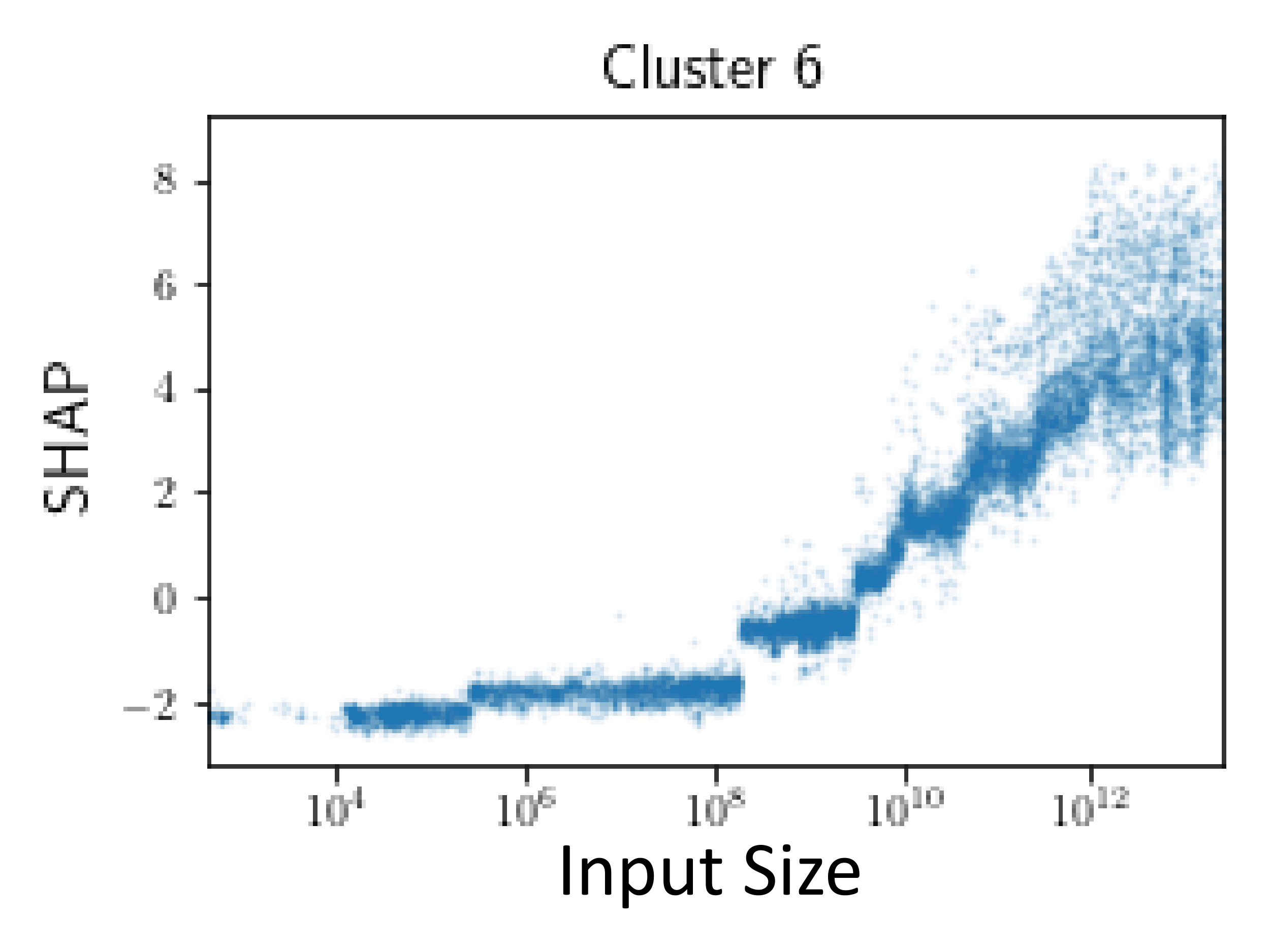}\vspace{-8pt}}
    \subfloat[By operator counts\label{fig:indi_contribution2}]{			\includegraphics[width=.4\columnwidth]{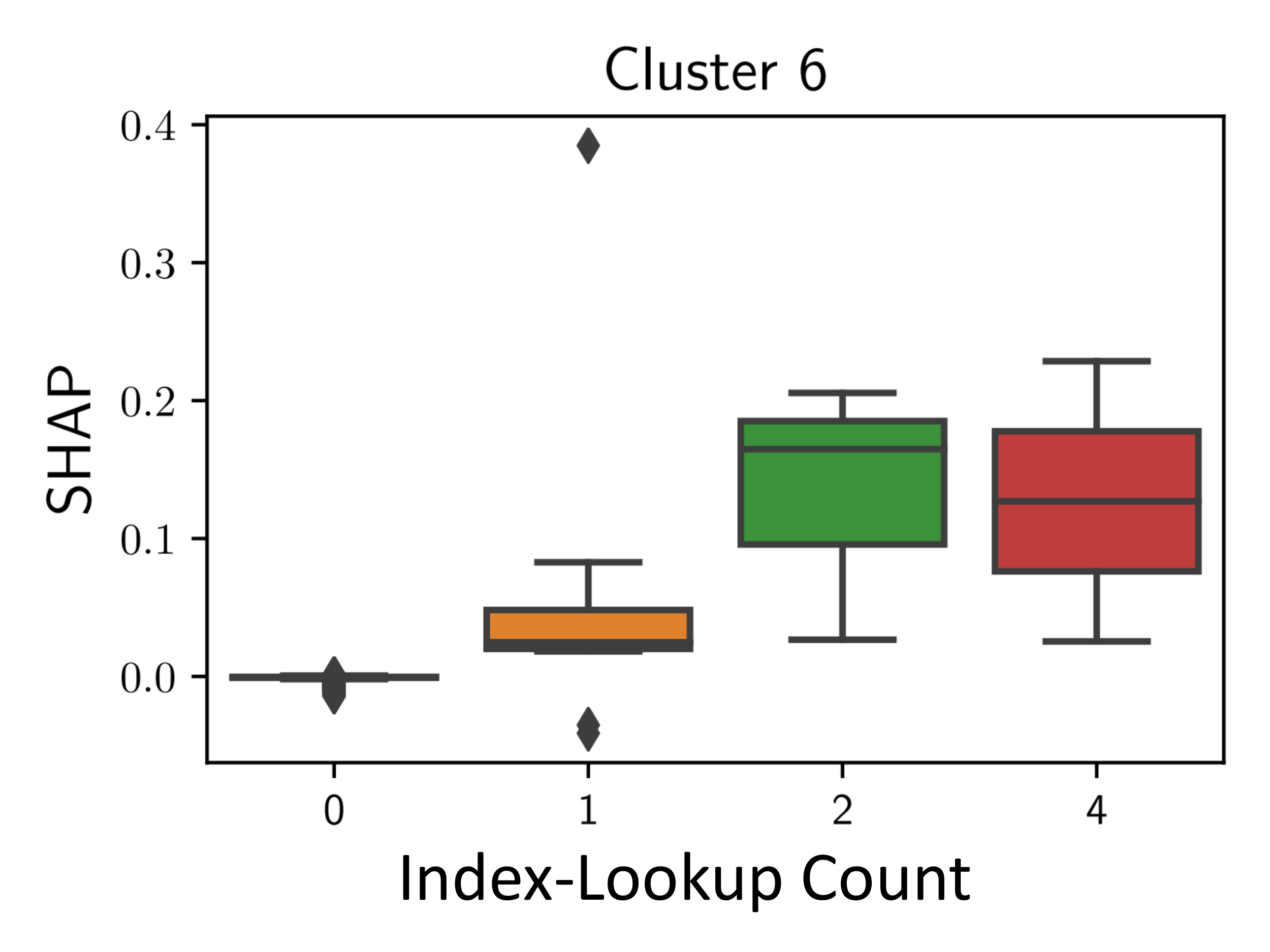}\vspace{-8pt}}
    \vspace{-2mm}
	\caption{SHAP value distribution.}
	\vspace{-0.6cm}
\end{figure} 
Figure~\ref{fig:indi_contribution} illustrates the distribution of Shapley values with respect to the total input data read, where each dot corresponds to one job instance. We can see that jobs with large input size are more likely to be in Cluster 6 (as their feature values lead to higher Shapley values and a thus higher likelihood of being in Cluster 6) using \deltanorm. Note that Cluster 6 has a relatively high variance and high probability of outliers. 
Similar trends can be found for jobs with fewer tokens.
\begin{center}
\shadowbox{
\begin{minipage}{0.9\linewidth}
    \textbf{Insight:} 
    Jobs with larger inputs and using fewer tokens are more likely to have a large variation. A larger number of tokens can potentially evacuate other jobs from the same machine, which potentially reduces interference and the impact of noisy neighbors.
\end{minipage}
}\vspace{-3pt}
\end{center}
Similarly, job characteristics such as operator counts significantly impact the prediction, indicating that the existence of certain operators more likely results in different runtime distributions (see Figure~\ref{fig:indi_contribution2} for Shapley values for Cluster 6 with \deltanorm).



\begin{center}
\shadowbox{
\begin{minipage}{0.9\linewidth}
    \textbf{Insight:} Certain operator counts, such as Index-Lookup, Window, and Range, increase the variation.
\end{minipage}
}\vspace{-3pt}
\end{center}


 
 

Using \rationorm, Cluster 0 has smaller variance and smaller probability of outliers than Cluster 2, while both have two modes. Focusing on a set of high-importance features, we compare the Shapley values for predicting these two clusters and found that, with lower CPU utilization, standard deviation and low usage of spare tokens, jobs are more likely to be in Cluster 0 (with more reliable performance) compared to Cluster 2. In general, we expect that machines with high utilization levels or standard deviations will have less reliable performance, which coincides with our observations here. The usage of spare tokens (whose availability is less predictable) can also lead to less stable runtimes. 

\begin{center}
\shadowbox{
\begin{minipage}{0.9\linewidth}
    \textbf{Insight:} 
    Lower CPU utilization (load), lower standard deviation, and less use of spare tokens can improve runtime reliability.
\end{minipage}
}\vspace{-3pt}
\end{center}

For \rationorm, increasing the vertex count on Gen5 and Gen6 (newer generations) tends to shift the prediction to Clusters 0 and 1, indicating that running vertices on those machine SKUs can potentially help with the runtime variation. Compared with Gen3 and Gen4 machines, those are in general faster and with large resource capacity~\cite{zhu2021kea}. 
 
 \begin{center}
\shadowbox{
\begin{minipage}{0.9\linewidth}
    \textbf{Insight:} The model identified certain SKUs where larger vertex count on those machines increases the likelihood of Clusters 0 and 1, which have smaller variance.
\end{minipage}
}\vspace{-3pt}
\end{center}

\section{Controlling Runtime Variation}
\label{sec:prescriptive}


Using the predictive model (Section~\ref{sec:model}) and drawing from the insights from the Shapley values (Section~\ref{sec:explain}), we now identify several what-if scenarios for scheduling and resource allocation and evaluate their performance. Based on the changes of jobs' runtime, one can quantitatively evaluate the detailed monetary impact.

\subsection{Scenario1: Spare Token allocation}

Availability of spare tokens depends on physical cluster conditions that are affected by the execution of other jobs and hence is a source of variation. Here we use our models to estimate the impact on runtime variation if spare tokens are not allocated.


 


With the predictor, we disable spare tokens for all jobs in the test set (dataset D3 as in Table~\ref{tab:datasets}). 
%
With \rationorm, 15\% of jobs that were predicted in Cluster 2 are now in Cluster 1, where the outlier probabilities gap between $25^{\text{th}}$ and $75^{\text{th}}$ percentiles, and the $95^{\text{th}}$ percentile of the normalized runtime are reduced (also see Table~\ref{tab:statscluster}). 
The second significant change is from Cluster 3 to 5 where the gap between 25 and $75^{\text{th}}$ percentile increased slightly from 0.17 to 0.19, while the standard deviation decreased dramatically (from 1.45 to 0.82). 
Similar changes can be seen for \deltanorm. Based on the predictor,
for the set of jobs desiring low variances, we propose disabling their usage of spare tokens to maximize performance improvement. \edit{In the production system, there has been work ongoing to reduce the maximum number for spare tokens as a multiplier of the number of allocated tokens. We observed that the jobs with fewer spare tokens runs slower but with less variance, which agrees with our model predictions.}



\subsection{Scenario2: Scheduling on later generation of machines}

A job's vertices can be executed by multiple machines in a distributed manner and as discussed in Section~\ref{subsec:sources-of-variation}, different job instances within the same group can be allocated to many different SKUs. Here we estimate the impact on runtime variation if we execute more vertices on later generations of machines.



By shifting all the vertices (both fractions and count) from Gen3.5 to Gen5.2,
%
%
the most likely change for 20.95\% of jobs is from Cluster 2 to 0, with a significant drop in the gap between 25 and $75^{\text{th}}$ percentile for \rationorm. And for \deltanorm, the most likely prediction change is from Cluster 1 to 0 where the gap between 25 and $75^{\text{th}}$ percentile dropped from 11s to 4s.



Hence, it's better to run more vertices on later generation SKUs. However, our model doesn't capture the compounding of changes due to workload re-balancing, such as the changes of CPU utilization levels.
Models that can predict the utilization levels given different workload distributions can be easily integrated, such as in KEA~\cite{zhu2021kea}, to quantitatively capture this dynamic impact.



\subsection{Scenario3: Improving load balance
}

As discussed in Section~\ref{subsec:sources-of-variation}, physical cluster conditions such as load differences across machines are a source of runtime variation. Here we estimate this impact of more uniformly distributed loads.




%
%
If the standard deviation of CPU utilization could be reduced to 0 (i.e., equal load on all machines and at all times), with \rationorm, the most likely change is from Cluster 2 to 0 (29.78\% of jobs), with outlier probability reduced from 1.66\% to 1.63\% and the variation measured by the difference between the $25^{\text{th}}$ and $75^{\text{th}}$ percentiles reduced from 0.16 to 0.06. Similar improvement can be seen for \deltanorm. 
Thus there is significant monetary value that can be realized by a better scheduling system that further motivates other research projects for \cosmos.




\section{Related Work}
\label{sec:related}


A number of works have studied the problem of predicting job runtimes under different resource allocations and platform parameter settings~\edit{\cite{pietri2014performance, tasq-arxiv, tasq-edbt, hu:reloca:infocom:2020, autodop, starfish:analytics-cluster-sizing:socc:2011, starfish:analytics-cluster-sizing:socc:2011, rajan2016perforator, guo2018machine}}.
Morpheus~\cite{jyothi2016morpheus} examines resource usages of recurring jobs and finds the best-fitting pattern that it uses for better resource allocation and scheduling, but does not predict the outlier probability for a given job. \edit{Runtime estimation has also been used to infer job scheduling to support policies, such as Shortest Processing time First (SPF)~\cite{tsafrir2007backfilling, gaussier2015improving, kuchnik2019ml, zotkin1999job}. Zrigui et al.~\cite{zrigui2022improving} clustered job instances into small and large based on runtime and used a classification algorithm to inform their job scheduler with high accuracy. 
In our work, we derived a larger number of clusters for delta- and ratio-normalized runtime distributions with richer information (percentiles, variances, etc.), providing a comprehensive oversight of customer experiences. 
}

Schad et al.~\cite{schad:ec2-variance:vldb:2010} studied performance variation on Amazon EC2~\cite{ec2} by running benchmarks for CPU and Memory performance, Disk I/O, and network bandwidth. 
Feitelson et al.~\cite{feitelson1995job} examined 
a parallel scientific workload on a 128-node cluster at NASA Ames and presented the changes of 
job submission rate, system utilization, and the distributions of job characteristics such as job type, runtime, and degree of parallelism. 
However, they do not model or predict the job runtime variation.

Prior works have proposed automated methods, including ML techniques, for analyzing system failures, slowdowns, and potential anomalies~\cite{bhaduri2011detecting,chitrakar2012anomaly,duan2009fa, shao2019griffon}. 
Causal inference and dependency/graph learning may also be used~\cite{heckerman1996causal,mandros2017discovering,zhang2020statistical,loh2014high,raskutti2018learning,nori2019interpretml,zheng2018hound} for these applications. However, in our work, where there is a large number of feature dimensions and complex correlation between groups of features, such methods require manual input to tune the dependency structure \edit{such as adding or deleting a detected dependency link} and might still be biased. \edit{In our work, we do not manually craft dependencies among causes of variation but use Shapley scores for inference.}









Huang et al.~\cite{huang2016identifying, huang2017top} investigated the causes for query latency variation in transactional databases.
They used variance (and covariance) as the primary metric for quantifying variation and built a variance tree corresponding to the call graph for gaining insights into the contributors of variance. In contrast, we use runtime \textit{distributions} for predicting variation, instead of scalar metrics due to the insufficiency of the latter in fully characterizing or explaining variation. Additionally, we use feature importance and Shapley scores to identify the main contributors to the prediction for the distribution to which a new run of a job may belong.

Prior work has also explored reducing performance variation with concurrent queries. Crescando~\cite{unterbrunner2009predictable} presents a relational table implementation that prioritizes predictable query performance 
over optimal performance 
through design choices such as having a scan-only architecture without indexes and new collaborative scan and update-join algorithms. CJOIN~\cite{candea2011predictable} introduces a new join operator and shares computation and resources among concurrent queries to improve both throughput and performance stability. Augmenting our prescriptive analysis with capabilities for evaluating the impact of computation sharing and other optimizations for concurrent queries is an interesting direction for future work. 

\edit{For tuning job performance, Black-box optimization such as Bayesian optimization~\cite{bishop2006pattern,curino2020mlos}, gradient descent~\cite{zhang2019end,van2017automatic}, etc. requires multiple runs of experiments. In this work, we provide a one-shot method that can directly determine the best course of action.
}
\section{Conclusion}
\label{sec:conclusion}

In this work, we did an extensive analysis of the runtime variation of recurring production jobs on Microsoft Cosmos by systematically characterizing, modeling, predicting, and explaining job runtime variations. Our original 2-step approach computes a posterior likelihood for each job to associate it with a predefined probability distribution, whose shape 
differs according to (1) intrinsic job characteristics, (2) resource allocation and (3) cluster condition when the job is submitted. 
We infer the distribution of job runtime with >96\% accuracy, out-performing the traditional regression models and capturing better the long tail of the distribution. 
Using an interpretable machine learning algorithm, we examined the sources of variation such as usage of spare tokens, skewed load on computing nodes, fractions of vertices executed on different SKUs. We quantified the improvement by adjusting these control variables. 
Our techniques can be used along with models that capture the effects on system utilization with workload re-balancing to dynamically optimize the performance of individual jobs.




\balance
\bibliographystyle{ACM-Reference-Format}
\bibliography{sample}

\end{document}